\documentclass[a4paper,12pt]{article}
\usepackage{mathrsfs}
\usepackage{amsmath,amssymb}
\usepackage{bm}
\usepackage{cite}
\def\beq{\begin{equation}}
\def\eeq{\end{equation}}
\def\bea{\begin{eqnarray}}
\def\eea{\end{eqnarray}}
\def\nn{\nonumber}

\def\ket#1{\left| #1\right\rangle}
\def\ket#1{\left| #1\right\rangle}
\def\N{{\cal N}}
\def\Sch#1{$\mathfrak{s}(#1)$}
\def\Q{\mathscr{Q}}
\def\S{\mathscr{S}}
\def\X{\mathscr{X}}
\def\SSch#1{$\mathfrak{s}(1/#1)$}
\def\del{\partial}
\newtheorem{prop}{Proposition}

\renewcommand{\labelenumi}{\roman{enumi})}
\begin{document}

\baselineskip=18pt

\thispagestyle{empty}

\vspace*{3cm}

\begin{center}
{\LARGE\sf
  Lowest Weight Representations of Super Schr\"odinger Algebras in 
  One Dimensional Space
}

\bigskip\bigskip
N. Aizawa

\bigskip
\textit{
Department of Mathematics and Information Sciences, \\
Graduate School of Science, 
Osaka Prefecture University, \\
Nakamozu Campus, Sakai, Osaka 599-8531, Japan}\\
\end{center}

\vfill
\begin{abstract}
 Lowest weight modules, in particular, Verma modules over the 
$ {\cal N} = 1, 2 $ super Schr\"odinger algebras in $ (1+1) $ dimensional spacetime are 
investigated. The reducibility of the Verma modules is analyzed via explicitly constructed 
singular vectors. The classification of the irreducible lowest weight modules is given for 
both massive and massless representations. A vector field realization of the $ {\cal N} = 1, 2 $ 
super Schr\"odinger algebras is also presented.
\end{abstract}

\clearpage
\setcounter{page}{1}
%
%
%
%
\section{Introduction}

  Conformal symmetry and conformal supersymmetry are one of the 
fundamental concept in relativistic field theories. 
However, importance of conformal invariance in nonrelativistic physics has been recognized since early 70's 
(see for example \cite{Nie,Hagen,Dunne,Henkel,Sachdev,MSW,NS} and references therein). 
Nonrelativistic counterpart of the conformal group is the Schr\"odinger group \cite{Nie,Hagen}. 
Recently, the Schr\"odinger group attracts very much attention in the context of 
nonrelativistic conformal field theory \cite{NS} and nonrelativistic version of 
AdS/CFT correspondence \cite{Son,BaMcG}. Supersymmetric extensions of the Schr\"odinger group and 
its Lie algebra have also been discussed in connection with various physical systems 
such as fermionic oscillator \cite{BH,BDH}, spinning particles \cite{GGT}, 
nonrelativistic Chern-Simons matter \cite{LLM,DH,NSY}, Dirac monopole and magnetic vortex \cite{DH} 
and so on. Some super Schr\"odinger algebras were constructed from the viewpoint of infinite dimensional 
Lie super algebra \cite{HU} or by embedding them to conformal superalgebras \cite{SY1, SY2}. 

  However, it seems that the representation theories of the Schr\"odinger group/algebra and 
their supersymmetric extensions are not studied well. 
There are a few works that were based on representation theoretic viewpoint. 
We mention the followings: 
Projective representations of the Schr\"odinger group in 3 spatial dimension were constructed in \cite{Perroud}. 
Irreducible representations of the Schr\"odinger algebras up to 3 spatial dimension were investigated in 
\cite{DDM,Mur,FKS}. The author of \cite{Naka} studied the highest weight representations of 
the $ {\cal N} =  2 $ super Schr\"odinger algebra 
in 2 spatial dimension ("exotic" algebra in the terminology of \cite{DH}).

  In a series of paper, we try to classify irreducible representations of the super Schr\"odinger algebras. 
This is motivated by the physical importance of the superalgebras mentioned above. 
We believe that the super Schr\"odinger algebra has a lot of potential applications in both classical and 
quantum physics. In the present paper, the first one of the series, 
we study the simplest case, that is, the super Schr\"odinger algebra 
in $(1+1)$ dimensional spacetime with $ {\cal N} = 1, 2 $ extensions. 
In general, the Schr\"odinger algebra of fixed spacetime dimension has some supersymmetric extension 
even for fixed value of $\cal N.$ A systematic method to extend the Schr\"odinger algebra in $(1+n)$ dimensional 
spacetime to arbitrary $ \cal N $ has developed in \cite{DH}. 
We study the super Schr\"odinger algebras introduced in \cite{DH} 
throughout the present paper. We denote the centrally extended Schr\"odinger algebra in $(1+n)$ dimensional 
spacetime by \Sch{n}. The $ {\cal N} = 1, 2 $ extension of \Sch{1} are denoted by \SSch{1}  and \SSch{2},
respectively. The superalgebra \SSch{2} corresponds to the one with $ N_+ = 1,\; N_- = 0 $ extension in \cite{DH}.

  The plan of this paper is as follows: We give the definitions of \SSch{1} and \SSch{2} 
in the next section. Two types of algebraic anti-automorphisms are also presented. 
In section \ref{LWR-N1} the lowest weight modules (Verma modules) of \SSch{1} are defined. 
To analyze the reducibility of the Verma modules we determine the singular vectors and 
give their explicit formula. Using the singular vectors the irreducible modules over \SSch{1} 
are classified for both massive and massless representations. 
The same analysis is repeated for \SSch{2} in section \ref{LWR-N2} and classification of 
the irreducible modules over \SSch{2} is presented. 
In section \ref{VFR} vector field realizations of \SSch{1} and \SSch{2} are introduced. 
Section \ref{ConRem} is devoted to concluding remarks.

%
%
%
\section{$ \N = 1, 2 $ extensions of \Sch{1}}

The algebra \Sch{1} has six elements, i.e., time translation $H,$ space translation $P,$ 
Galilei boost $G,$ dilatation $D,$ conformal transformation $K$ and central element $M$ 
corresponding to the mass.  
The nonvanishing commutation relations of \Sch{1} are given by:
\bea
        \begin{array}{llll}
          [H, D] = 2H, & \quad [H, K]=D, & \quad [D, K ]= 2K, & \quad [P, G] = M, \\[3pt]
          [H, G] = P,  & \quad [D, G] = G, & \quad [P, D] = P, & \quad [P, K] = G.
        \end{array}
        \label{S1bosonic}
\eea 
$ \N = 1 $ extension of \Sch{1} is defined by adding three odd elements  
$ \Q, \S, \X $ to \Sch{1}. They enjoy the anti-commutation relations given below
\bea
  \begin{array}{lll}
      \{ \Q, \Q \} = -2H, & \quad \{ \S, \S \} = -2K, & \quad \{ \X, \X \} = -M, \\[3pt]
      \{ \Q, \X \} = -P, & \quad \{ \S, \X \} = -G, & \quad \{ \Q, \S \} = -D,
  \end{array}
  \label{S1N1fermionic}
\eea
and satisfy the nontrivial commutation relations with \Sch{1}:
\beq
  \begin{array}{llll}
    [\Q, D] = \Q, & \quad [\Q, K] = \S, &  \quad [D, \S] = \S, &  \quad [H, \S] = \Q, \\[3pt]
    [\Q, G] = \X, &  \quad [P, \S] = \X.
  \end{array}
  \label{S1N1BS}
\eeq
The subset $ \{ H, D, K, T, Q \} $ forms a subalgebra of \SSch{1} isomorphic to $ osp(1/2). $

$ \N = 2 $ extension of \Sch{1} has six odd elements $ \Q_i, \S_i, \X_i \; (i=1,2). $ 
In addition to these, there exist another even element, denoted by $R_{12},$ which commutes 
with all the generators of \Sch{1} so that $ \N = 2 $ extended \Sch{1} has 
13 generators. 
Nontrivial relations of the generators newly added are listed below. 
The anti-commutators of odd elements:
\beq
  \begin{array}{lll}
    \{ \Q_j, \Q_k \} = -2 \delta_{jk} H, & \quad \{ \S_j, \S_k \} = -2 \delta_{jk} K & \quad 
    \{ \X_j, \X_k \} = - \delta_{jk} M, 
    \\[3pt]
    \{ \Q_j, \X_k \} = - \delta_{jk} P, & \quad \{ \S_j, \X_k \} = - \delta_{jk} G, & \quad
    \{ \Q_j, \S_k \} = - \delta_{jk} D + R_{jk},
  \end{array}
  \label{S1N2fermionic}
\eeq
and the commutators between odd and even elements:
\beq
  \begin{array}{llll}
     [ \Q_j, D ] = \Q_j, & \quad [ \Q_j, K ] = \S_j, & \quad [ D, \S_j ] = \S_j, & \quad [ H, \S_j ] = \Q_j, \\[3pt]
     [ \Q_j, G ] = \X_j, & \quad [ P, \S_j ] = \X_j, & 
     \multicolumn{2}{l}{\quad [ \Q_j, R_{k \ell} ] = \delta_{j\ell} \Q_k - \delta_{jk} \Q_{\ell},} \\[3pt]
     [ \S_j, R_{k \ell} ] = \delta_{j\ell} \S_k - \delta_{jk} \S_{\ell}, 
     & \multicolumn{3}{l}{\quad  [ \X_j, R_{k \ell} ] = \delta_{j\ell} \X_k - \delta_{jk} \X_{\ell},}
  \end{array}
  \label{S1N2BS}
\eeq
where $ j, k \in \{ 1, 2 \} $ and we impose $ R_{ij} = - R_{ji}. $

  The defining relations of \SSch{2} given above are not appropriate to study 
lowest or highest weight representations. We thus take a  linear combination of the odd generators:
  \beq
    R = i R_{12}, \quad \mathscr{A}_{\pm} = \frac{1}{ \sqrt{2} } (\mathscr{A}_1 \pm i \mathscr{A}_2),
    \label{S1N2combination}
  \eeq
where $ \mathscr{A}_j = \Q_j, \S_j, \X_j. $ 
Then the elements $ \mathscr{A}_{\pm} $ enjoy the fermionic property: $ \mathscr{A}_{\pm}^2 = 0. $ 
With the new generators, the counterpart of the relations (\ref{S1N2fermionic}) reads
\beq
  \begin{array}{lll}
     \{ \Q_{\pm}, \Q_{\mp} \} = - 2H, & \quad \{ \S_{\pm}, \S_{\mp} \} = - 2K, & \quad \{ \X_{\pm}, \X_{\mp} \} = -M,
     \\[3pt]
     \{ \Q_{\pm}, \X_{\mp} \} = -P, & \quad \{ \S_{\pm}, \X_{\mp} \} = -G, &  \quad 
     \{ \Q_{\pm}, \S_{\mp} \} = - D \mp R,
  \end{array}
  \label{S1N2fermionic2}
\eeq
and that of (\ref{S1N2BS}) is given by
\beq
  \begin{array}{llll}
    [ \Q_{\pm}, D ] = \Q_{\pm}, & \quad [ \Q_{\pm}, K ] = \S_{\pm}, &  \quad [D, \S_{\pm} ] = \S_{\pm}, 
    & \quad [ H, \S_{\pm} ] = \Q_{\pm} 
    \\[3pt]
    [ \Q_{\pm}, G ] = \X_{\pm}, &  \quad [ P, \S_{\pm} ] = \X_{\pm}, & \quad [ R, \mathscr{A}_{\pm}] = \pm \mathscr{A}_{\pm}.
  \end{array}
  \label{S1N2BS2}
\eeq
The subset $ \{ H, D, K, R, \Q_{\pm}, \S_{\pm} \} $ forms a subalgebra of \SSch{2} isomorphic to $ osp(2/2). $

  One can define two types of adjoint operations (algebra anti-automorphism)\cite{SNR} to 
the super Schr\"odinger algebras introduced above. 
The first one maps even elements as follows:
\beq
  \omega_1(P) = (-1)^{\epsilon} G, \quad \omega_1(H) = K, \quad \omega_1(D) = D, \quad \omega_1(M) = M, 
  \quad \omega_1(R) = R,
  \label{omega1-even}
\eeq
where $ \epsilon \in \{ \ 0, 1 \ \}. $ 
The mapping of the odd elements of \SSch{1} is given by
\beq
  \omega_1(\Q) = (-1)^{\lambda} \S, \quad \omega_1(\X) = (-1)^{\epsilon+\lambda} \X,
  \label{omega1-odd1}
\eeq
and of \SSch{2} by
\beq
  \omega_1(\Q_a) = (-1)^{\lambda} \S_{-a}, \quad \omega_1(\X_a) = (-1)^{\epsilon+\lambda} \X_{-a},
  \label{omega1-odd2}
\eeq
where $ \lambda \in \{ \ 0, 1 \ \} $ and $ a = \pm. $ 
The second one is defined by
\beq
  \omega_2(P) = (-1)^{\epsilon} P, \quad \omega_2(G) = (-1)^{\epsilon} G, \quad
  \omega_2(X) = -X \ (X = H, K, D, M), \quad \omega_2(R) = R,
  \label{omega2-even}
\eeq
together with for $ \N = 1 $
\beq
  \omega_2(\Q) = i(-1)^{\lambda} \Q, \quad \omega_2(\S) = i(-1)^{\lambda} \S, 
  \quad
  \omega_2(\X) = i(-1)^{\lambda+\epsilon+1} \X,
  \label{omega2-odd1}
\eeq
and for $ \N = 2$
\beq
  \omega_2(\Q_a) = i(-1)^{\lambda} Q_{-a}, \quad
  \omega_2(\S_a) = i(-1)^{\lambda} \S_{-a}, \quad
  \omega_2(\X_a) = i(-1)^{\lambda+\epsilon+1} \X_{-a}.
  \label{omega2-odd2}  
\eeq
Both adjoint operations are idempotent: $ \omega_j^2 = id. $ 

  On the contrary, there is no grade adjoint operation (another algebra anti-automorphism) \cite{SNR} 
for \SSch{1} although \SSch{1} has $ osp(1/2) $ as a subalgebra. 
To see this fact, we note that the grade adjoint operation $ \sigma $ preserves the parity of 
an element. For instance, $ \sigma(\Q) $ is a linear combination of $ \Q, \S $ and $ \X. $ 
We thus write the mapping in a matrix form:
\[
   \sigma
   \begin{pmatrix}
     \Q \\ \S \\ \X
   \end{pmatrix}
   = 
   A
   \begin{pmatrix}
     \Q \\ \S \\ \X
   \end{pmatrix},
\]
where $A$ is a $ 3 \times 3 $ complex matrix. 
By the definition of the grade adjoint operation $ \sigma^2({\mathscr A}) = - \mathscr{A} $ 
for an odd element $ \mathscr{A}. $ 
Therefore the matrix $ A $ has to satisfy 
$ \bar{A} A = -1_3. $ This yields a contradictory result: $ | \mbox{det} A|^2 = -1. $ 

  The superalgebra \SSch{2} admits two types of grade adjoint operations. 
The first one is given by
\bea
 &  & 
  \sigma_1(\Q_{\pm}) = \pm(-1)^{\epsilon} \S_{\mp}, \quad
  \sigma_1(\X_{\pm}) = \pm(-1)^{\epsilon} \X_{\mp}, \quad
  \sigma_1(K) = H, \quad \sigma_1(P) = G, 
 \nn \\
 & &
  \sigma_1(X) = X \ ( X= D, R, M),
  \label{GAO1-S1}
\eea
where $ \epsilon \in \{ \ 0, 1 \ \}. $
The second one looks as follows:
\beq
  \sigma_2( {\mathscr A}_{\pm}) = \pm(-1)^{\epsilon} i {\mathscr A}_{\mp},
  \quad 
  \sigma_2(R) = R, \quad 
  \sigma_2(X) = -X \ ( X = H, K, D, M, P, G).
  \label{GAO2-S1}
\eeq

%
%
%
\setcounter{equation}{0}
\section{Lowest weight representations of \bm{\SSch{1}}}
\label{LWR-N1}

\subsection{Verma modules and singular vectors}

  The superalgebra \SSch{1} is graded, in addition to the $ {\mathbb Z}_2 $ grading of 
Lie superalgebra, if we define
\bea
 & & 
 \mbox{deg} K = 2, \quad \mbox{deg} G = \mbox{deg} \S = 1, \quad 
 \mbox{deg} D = \mbox{deg} M = \mbox{deg} \X = 0, \nn
 \\
 & &
 \mbox{deg} P = \mbox{deg} \Q = -1, \quad \mbox{deg} H = -2. 
 \label{degSS1}
\eea
The grading operator is $D$. The $ \mathbb Z $ grading defined above can be viewed as an 
analogue of the triangular decomposition of semisimple Lie algebra. 
Namely, one has the vector space decomposition defined by  
\SSch{1} $= {\mathfrak s}(1/1)^+ \oplus {\mathfrak s}(1/1)^0 \oplus {\mathfrak s}(1/1)^- $ 
where $ {\mathfrak s}(1/1)^+ ({\mathfrak s}(1/1)^-) $ is spanned by the generators of positive (negative)
degree and $ {\mathfrak s}(1/1)^0 $ is by zero degree. 
We remark that each subset $ {\mathfrak s}(1/1)^{\pm}, {\mathfrak s}(1/1)^0 $ 
forms an abelian subalgebra. 
We also remark that the adjoint operation $ \omega_1$ introduced in the previous section 
exchanges $ {\mathfrak s}(1/1)^+ $ and $ {\mathfrak s}(1/1)^-,$ while $ \omega_2 $ preserves the 
triangular structure of \SSch{1}. 

  The triangular decomposition enables us to define lowest weight modules of \SSch{1}, in 
particular, Verma modules. 
The lowest weight vector $ v_0 $ is defined by
\bea
 & & \Q \, v_0 = P \, v_0 = 0, \nn \\
 & & D v_0 = -d v_0, \quad M v_0 = m v_0, \quad  \X v_0 = \chi v_0,
 \label{}
\eea
where $d \in {\mathbb R} $ is the conformal weight and the minus sign is for later 
convenience. The variable $\chi$ is of odd parity relating to the mass eigenvalue 
by the relation $ m=2 \chi^2. $ 
The Verma module $ V^d $ is defined as the lowest weight module with the lowest weight $d: $
$ V^d = U({\mathfrak s}(1/1)^+) \otimes v_0, $ where $ U({\mathfrak s}(1/1)^+) $ denotes 
the universal enveloping algebra of $ {\mathfrak s}(1/1)^+.$ 
More explicitly, 
\beq
  V^d = \{  \ G^k K^{\ell} v_0, \ G^k K^{\ell} \S v_0 \ \}, 
  \label{VdS1}
\eeq
where $ k, \ell $ are non-negative integers. 
Note that the subset $ \{  \ G^k K^{\ell} v_0 \ \} $ is a Verma module of the non-super Schr\"odinger 
algebra \Sch{1}. 
It is an easy exercise to compute the action of \SSch{1} on the basis of $ V^d. $ 
Denoting $ v_{k,\ell} = G^k K^{\ell} v_0 $ and $ \nu_{k, \ell} =  G^k K^{\ell} \S v_0, $ 
one finds the followings:
\bea
  & & K  v_{k,\ell} = v_{k,\ell+1}, \qquad K \nu_{k,\ell} = \nu_{k,\ell+1}, \qquad
    G \, v_{k,\ell} = v_{k+1,\ell}, \qquad G \, \nu_{k,\ell} = \nu_{k+1,\ell},
  \nn \\
  & & D \, v_{k,\ell} = (k+2\ell-d) \, v_{k,\ell}, \qquad D \, \nu_{k,\ell} = (k+2\ell+1-d) \, \nu_{k,\ell},
  \label{ActionS1onVd} \\
  & & M v_{k,\ell} = mv_{k,\ell}, \qquad M \nu_{k,\ell} = m \nu_{k,\ell}, \qquad
      \X v_{k,\ell} = \chi \, v_{k,\ell}, \qquad \X \nu_{k,\ell} = \chi \, \nu_{k,\ell} - v_{k+1,\ell},
  \nn \\
  & & P \,v_{k,\ell} = \ell v_{k+1,\ell-1} + mk \,v_{k-1,\ell}, \qquad
      P \, \nu_{k,\ell} = \ell \nu_{k+1,\ell-1} + \chi v_{k,\ell} + mk\, \nu_{k-1,\ell},
  \nn \\
  & & Q \, v_{k,\ell} = \chi k \, v_{k-1,\ell} + \ell \nu_{k,\ell-1}, \qquad  
      Q \, \nu_{k,\ell} = \chi k \, \nu_{k-1,\ell} + (d-\ell-k) v_{k,\ell},
  \nn \\
  & & H v_{k,\ell} = \ell (k+\ell-d-1) v_{k,\ell-1} + \frac{1}{2} mk (k-1) v_{k-2,\ell},
  \nn \\
  & & H \nu_{k,\ell} = \ell (k+\ell-d) \nu_{k,\ell-1} + \chi k \, v_{k-1,\ell} + \frac{1}{2} mk(k-1) \nu_{k-2,\ell}.
  \nn
\eea
It follows that the Verma module $ V^d $ can be decomposed into homogeneous subspaces with respect to $D$:
\beq
   V^d = \bigoplus_{n=0}^{\infty} V_n^d, \qquad
   V_n^d = \mbox{lin.span.} \{ v_{k,\ell}, \ \nu_{k, \ell} \ | \ D  v_{k,\ell} = n v_{k,\ell}, \ D \nu_{k, \ell} = n \nu_{k, \ell} \}.
   \label{HomoDecompVd}
\eeq

   To analyse the reducibility of $ V^d, $ one can use the singular vectors\cite{Naka,DDM,Mur} 
although the superalgebra \SSch{1} is not semisimple. 
A singular vector $v_s$ is defined as a homogeneous element of $ V^d $ such that $ v_s \neq {\mathbb C} v_0 $ and 
\beq
   \Q \, v_s = P \, v_s = 0.  \label{SVS1}
\eeq
We give all the possible singular vectors explicitly. To this end, 
the following proposition on non-super case is helpful \cite{DDM,Mur}.
\begin{prop} \label{prop1}
 The singular vectors of the Verma module over \Sch{1} are given as follows:
 \begin{enumerate}
    \item For $ m \neq 0, $ a singular vector exists for $ d = p-3/2 $ which is given by 
      $
        ( G^2 - 2mk)^p w_0.
      $
    \item For $ m = 0, $ infinitely many singular vectors exist for each value of $d$ which are given by 
      $ G^p w_0. $
 \end{enumerate}
 In both cases, $ p \in {\mathbb N} $ and $ w_0 $ denotes a lowest weight vector of \Sch{1} defined by 
 $ H w_0 = P w_0 =0. $
\end{prop}

 Since a singular vector of \SSch{1} is a homogeneous element in $ V^d,$ it may have the form of 
\beq
  v_s = f(G,K) u_0, \quad u_0 = ( \alpha G + \beta \S) v_0, \label{SVassumpS1}
\eeq
where $ f(G,K) $ is a homogeneous polynomial in $ G, K $ and $ \alpha, \beta $ are constant of party even and odd, 
respectively. 
Assuming the degree of $ f(G,K) $ is $n$, the general expression of $ f(G,K) $ is
\beq
  f(G,K) = \sum_{\ell} a_{\ell} G^{n-2\ell} K^{\ell}. \label{fofGK}
\eeq
The action of $ P$ on $ v_s$ is computed as follows:
\bea
  P v_s &=& ([P, f(G,K)] + f(G,K) P) u_0
  \nn \\
  &=& 
   \sum_{\ell} \{ m(n-2\ell) a_{\ell} + (\ell+1) a_{\ell+1} \} \, G^{n-2\ell-1} K^{\ell} u_0
   + f(G,K)P u_0.
   \nn
\eea
Since $ P u_0 = ( \alpha m + \beta \chi) v_0, $ $ P v_s = 0 $ implies that $ P u_0 = 0. $ 
Clearly, $ H u_0 = 0 $ and $ D u_0 = -(d-1) u_0. $ Thus the vector $ u_0 $ is a lowest weight vector of the 
non-super Schr\"odinger algebra \Sch{1} with the conformal weight $ d-1. $ 

  One can determine the singular vectors of \SSch{1} applying Proposition \ref{prop1}.  
If $ m \neq 0, $ then $ P u_0 = 0 $ yields $ \beta = -2 \alpha \chi.$ 
One see from Proposition \ref{prop1} that if $ d = p-1/2,$ then 
$ v_s^p = (G^2-2mK)^p (G - 2 \chi \S) v_0 $ is the only vector annihilated by $P$ and $ H. $
It is verified by direct computation that $ Q v_s^p = 0. $ 
Thus $ v_s^p $ is a singular vector of \SSch{1} for $ m \neq 0. $ 
If $ m = 0, $ then $ \chi $ becomes a Grassmann number $(\chi^2 = 0) $ and $ P u_0 = 0 $ yields $ \beta = 0. $ 
In this case, there are infinite number of vectors $ v_s^p = G^p v_0 \ (p \in {\mathbb N}) $ for 
each value of $d.$ It is straightforward to see that $ \Q\, v_s^p =0. $ 
Therefore we have proved the following proposition.
\begin{prop} \label{prop2}
 The singular vectors of the Verma module $ V^d $ over \SSch{1} are given as follows: 
 \begin{enumerate}
   \item For $ m \neq 0, $ a singular vector exists for $ d = p - 1/2 $ which is
     \beq
       v_s^p = (G^2-2mK)^p (G-2\chi \S) v_0, \quad p \in {\mathbb Z}_{\geq 0}  \label{SVN1}
     \eeq
   \item For $ m = 0, $ infinitely many singular vector exist for each value of $d$ which are
     \beq
       v_s^p =  G^p v_0, \quad p  \in {\mathbb N} \label{SVN1m0}
     \eeq
 \end{enumerate}
\end{prop}
We remark that the case of $m=0$ corresponds to the super Schr\"odinger algebra 
without central extension. Proposition \ref{prop2} tells that the representations of the centrally 
extended super Schr\"odinger algebra is very different from the unextended one as in the non-super 
algebra \cite{DDM, Mur}.

\subsection{Irreducible modules}
\label{IMN1}

 Let us consider the reducibility of the Verma modules $V^d. $ We start with $ m \neq 0. $  
If $ d \neq p-1/2, $ there is no singular vectors in $ V^d $ so that the module is 
irreducible. In the case of $ d = p-1/2, $ there is one singular vector (\ref{SVN1}) in $ V^d. $
The submodule $ I^d = U({\mathfrak s}(1/1)^+) \otimes v_s^p $ is invariant under the action of 
\SSch{1}. This means that the Verma module $ V^d $ with $ d = p-1/2 $ is not irreducible. 
Furthermore, the submodule $ I^d $ is isomorphic to $ V^{d'} $ with shifted weight 
$ d' = d - 2p-1 = -p-3/2. $ This Verma module $ V^{d'} $ with the shifted weight does not 
have singular vectors since its weight has upper bound $ d' < -3/2, $ while Proposition \ref{prop2} 
shows that the Verma module has a singular vector provided that its weight is $ d > -1/2. $    

  We now consider the factor module $ V^d/I^d $ with $ d = p-1/2.$ 
Its lowest weight vector is denoted by $ w_0^p. $ It satisfies the conditions:
\beq
  Q w_0^p = P w_0^p = (G^2-2mK)^p (G - 2 \chi \S) w_0^p = 0. \label{LWfactor}
\eeq
The vectors $ G^k K^{\ell} w_0^p, \ G^k K^{\ell}\S w_0^p $ 
with $ k, \ell $  non-negative integers span $ V^d/I^d. $  
By the relation obtained from  (\ref{LWfactor}) one can get rid of all powers of 
$ G $ greater than $2p$:
\beq
  G^{2p+1} w_0 = 2 \chi G^{2p} \S w_0^p 
  - \sum_{j=0}^{p-1} \left( \begin{array}{c} p \\ j \end{array} \right) (-2m)^{p-j}
    G^{2j} K^{p-j} (G - 2\chi \S) w_0^p.
    \label{G2p+1}
\eeq
Hence the basis of $ V^d/I^d $ is given by 
\beq
  w_{k,\ell} =  G^k K^{\ell} w_0^p, \quad \omega_{k,\ell} = G^k K^{\ell}\S w_0^p, \quad k \leq 2p
  \label{FactorBasis}
\eeq
Search for the singular vectors in $ V^d/I^d $ is straightforward but requires lengthy computation. 
We omit the computation but the result is that there are no singular vectors in $ V^d/I^d. $ 
We thus conclude that $ V^d/I^d $ is irreducible. 

 Next we investigate the case of $ m = 0. $ All Verma modules in this case are reducible, 
since they contain infinitely many singular vectors. For a fixed value of $d, $ the Verma module 
contains infinite submodules 
$ I^p = U({\mathfrak s}(1/1)^+) \otimes v_s^p $
which are invariant under the action of \SSch{1}. 
Since the singular vector  $ v_s^p \ (p > 1)$ is related to $ v_s^1 $ by the relation 
$ v_s^p = G^{p-1} v_s^1, $ the corresponding submodules satisfy the inclusion relation  
$ I^p \subset I^1.$ It follows that
\[
  V^d / I^1 \subset V^d / I^p.
\] 
Therefore it is sufficient to examine the factor module $ V^d / I^1. $ 
We denote the lowest weight vector of  $ V^d / I^1 $ by $ w_0. $ It satisfies the conditions:
\beq
  Q w_0 = P w_0 = G w_0 = 0. \label{LWfactorm0}
\eeq
The basis of $ V^d / I^1 $ is given by
\[
  w_{\ell} = K^{\ell} w_0, \quad \omega_{\ell} = K^{\ell} \S w_0. \label{FactorBasism0}
\]
It is easy to see the action of $\Q$ and $ P $ on the basis:
\beq
  \Q w_{\ell} = \ell \omega_{\ell-1}, \quad \Q \omega_{\ell} = (d - \ell) w_{\ell}, \quad
  P w_{\ell} = P \omega_{\ell} = 0.
  \label{QPonW}
\eeq
It follows that if $ d \notin {\mathbb N} $ there is no singular vector in $ V^d / I^1, $
therefore the module $ V^d / I^1 $ is irreducible. If $ d = p\in {\mathbb N} $ there is one 
singular vector: $ w_s^p = \omega_p. $ 
It produces the submodule $ {\cal I}^p = U({\mathfrak s}(1/1)^+) \otimes w_s^p $ invariant under 
the action of \SSch{1}. The lowest weight vector $ \ket{0} $ of the factor module $ (V^p / I^1)/{\cal I}^p $ 
satisfies the condition
\beq
   Q \ket{0} = P \ket{0} = G \ket{0} = K^p \S \ket{0} = 0. \label{LWket0}
\eeq
The basis of $ (V^p / I^1)/{\cal I}^p $ is given by 
\[
  K^{\ell} \ket{0}, \quad K^{\ell} \S \ket{0} \ ( \ell = 0, 1, \cdots, p-1), \quad K^p \ket{0}.
\]
It is clear that there is no singular vector in the factor module $ (V^p / I^1)/{\cal I}^p $ 
thus the module is irreducible and is of dimension $ 2p+1. $ 
The generators $ P, G, M $ and $ \X $ are represented trivially in $ V^d / I^1 $ and $ (V^p / I^1)/{\cal I}^p $ 
if we set $ \chi = 0. $ One may identify $ V^d / I^1 $ and $ (V^p / I^1)/{\cal I}^p $ as infinite and finite 
dimensional representations of the  subalgebra $ osp(1/2). $ 
Now we summarize our results in the following proposition. 
\begin{prop} \label{prop3}
  The irreducible lowest weight modules over the Lie superalgebra \SSch{1} are classified as follows:
  \begin{enumerate}
   \item $ m \neq 0$
      \begin{itemize}
         \item $ V^d  $ when $ d \neq p-1/2, \ p \in {\mathbb Z}_{\geq 0}.$ dim $ V^d = \infty$
         \item $ V^d / I^d $ when $ d = p -1/2, \ p \in {\mathbb Z}_{\geq 0}. $ dim $ V^d / I^d = \infty $
      \end{itemize}
   \item $ m = 0$
      \begin{itemize}
         \item $ V^d / I^1 $ when $ d \notin {\mathbb N}.$ dim $ V^d / I^1 = \infty$
         \item $ (V^p / I^1)/{\cal I}^p $ when $ p \in {\mathbb N}.$ dim $ (V^p / I^1)/{\cal I}^p = 2p+1$
      \end{itemize}
      The representations for $ m = 0 $ are also irreps of the subalgebra $ osp(1/2). $ 
  \end{enumerate}
\end{prop}

%
%
%
\setcounter{equation}{0}
\section{Lowest weight representations of \bm{\SSch{2}}}
\label{LWR-N2}

\subsection{Verma modules and singular vectors}

 The Lie superalgebra \SSch{2} has $ {\mathbb Z} \times {\mathbb Z} $ grading structure 
if we define
\bea
  & & \mbox{deg} K = (2,0), \qquad \mbox{deg} G = (1,0), \qquad \mbox{deg} D = \mbox{deg}  R = \mbox{deg}  M  = (0,0),
  \nn \\
  & & \mbox{deg} H = (-2,0), \qquad \mbox{deg} P = (-1,0),  \qquad 
   \label{degSS2} \\
  & & \mbox{deg} \S_{\pm} = (1, \pm 1), \qquad \mbox{deg}\X_{\pm} = (0,\pm 1), \qquad \mbox{deg} \Q_{\pm} = (-1,\pm 1).
   \nn
\eea
The grading operators are $ D $ and $R.$ 
The grading enable us to introduce a triangular decomposition of \SSch{2}. 
Reading the degree of an element from left to right 
if the first encountered non-zero entry is positive (negative), then we say that the element has 
positive (negative) degree.   
We define the following vector space decomposition according to the degree of elements:
\bea
  & & {\mathfrak s}(1/2) = {\mathfrak s}(1/2)^+ \oplus {\mathfrak s}(1/2)^0 \oplus {\mathfrak s}(1/2)^-,
  \nn \\
  & & {\mathfrak s}(1/2)^+ = \{ \ K, G, \S_{\pm}, \X_+ \ \}, \qquad
      {\mathfrak s}(1/2)^0 = \{ \ D, R, M \ \},
  \nn \\
  & & {\mathfrak s}(1/2)^- = \{ \ H, P, \Q_{\pm}, \X_- \ \}.
  \nn
\eea
In contrast to \SSch{1}, the subsets $ {\mathfrak s}(1/2)^{\pm} $ form subalgebras but they are not abelian. 

  We define the lowest weight vector $ v_0 $  by
\bea
  & & \Q_{\pm} v_0 = P v_0 = \X_- v_0 = 0,
  \label{LWS2} \\
  & & D v_0 = -d v_0, \quad M v_0 = m v_0, \quad R v_0 = r v_0.
  \nn
\eea
The Verma modules over \SSch{2} are defined by
\beq
  V^{d,r} = \{ \ G^k K^{\ell} \S_+^a \S_-^b \X_+^c v_0 \ | \ k, \ell \in {\mathbb Z}_{\geq 0}, a, b, c \in \{ 0, 1 \} \ \}.
  \label{VMS2}
\eeq
The action of the grading operators on the basis $ v_{k,\ell,a,b,c} = G^k K^{\ell} \S_+^a \S_-^b \X_+^c v_0 $ 
is easily computed:
\beq
  D \, v_{k,\ell,a,b,c} = (k + 2\ell+a+b-d) v_{k,\ell,a,b,c}, \qquad R \, v_{k,\ell,a,b,c} = (a-b+c+r) v_{k,\ell,a,b,c}.
  \label{DRonbasis}
\eeq
It follows that the Verma module $ V^{d,r} $ can be decomposed into homogeneous subspaces with respect to $D$ and $R$:
\bea
 &&
   V^{d,r} = \bigoplus_{n_1,n_2=0}^{\infty} V_{n_1,n_2}^{d,r}, 
   \label{HomoDecompVd2} \\
 &&
   V_{n_1,n_2}^{d,r} = 
   \mbox{lin.span.} \{ v_{k,\ell,a,b,c}, \ | \ D  v_{k,\ell,a,b,c} = n_1 v_{k,\ell,a,b,c}, \ R v_{k,\ell,a,b,c} = n_2 v_{k,\ell,a,b,c} \}.
   \nn
\eea

Since a singular vector of \SSch{2} is a homogeneous element in $ V^{d,r}$ 
and $ \S_{\pm}, \X_+ $ are nilpotent, it may have the factorized form:
\beq
  v_s = f(G,K) u_0,  \label{SVassumpS2}
\eeq
where $ f(G,K) $ is a homogeneous polynomial in $ G, K $ and $ u_0 $ is a homogeneous 
element of $ V^{d,r} $ containing fermionic generators. 
We note that $ \X_- $ commutes with $ f(G,K).$ This gives a necessary condition for $ u_0:$
\beq
   \X_- u_0 = 0. \label{condition1}
\eeq 
There are eight possible $ u_0 $ according to the values of $ a, b $ and $ c. $ 
It is seen from (\ref{DRonbasis}) that 
they are specified by the pair of integers $(a+b, a-b+c). $ 
We call this degree of $ u_0. $ All possible $ u_0 $ are listed in Table~1. 
\begin{table}[h]
     \renewcommand{\arraystretch}{1.2}
     \renewcommand{\tabcolsep}{3mm}
 \begin{center}
  \begin{tabular}{ccccccc}
    \hline
       & $a$ & $b$ & $c$ & $ (a+b, a-b+c) $ & $ u_0 $ \\
    \hline\hline
      i) & 0 & 0 & 0 & (0, 0) & $v_0$ \\
      ii) & 0 & 0 & 1 & (0, 1) & $ \X_+ v_0 $ \\
      iii) & 0 & 1 & 0 & (1, -1) & $ \S_- v_0 $ \\
      iv) & 1 & 0 & 0 & (1, 1) & $ (\alpha \S_+ + \beta G \X_+) v_0 $ \\
      v) & 0 & 1 & 1 & (1, 0) & $ ( \alpha \S_- \X_+ + \beta G) v_0 $ \\
      vi) & 1 & 0 & 1 & (1, 2) & $ \S_+ \X_+ v_0 $ \\
      vii) & 1 & 1 & 0 & (2, 0) & see text \\
      viii) & 1 & 1 & 1 & (2, 1) & see text \\
    \hline
  \end{tabular}
 \end{center}
 \caption{List of $ u_0 $ and its degree}
\end{table}
We examine the eight cases to seek possible singular vectors. 

\begin{enumerate}
\renewcommand{\labelenumi}{\roman{enumi})}
\item 
deg $ u_0 = (0,0) :$ Since $ P v_0 = H v_0 = 0, $ $ u_0 $ is the lowest weight vector of non-super \Sch{1}. 
From Proposition \ref{prop1} a candidate of the singular vector for $ m \neq 0 $ is given by 
$ v_s = (G^2 - 2mK)^p v_0 $  and for $ m = 0 $ by $ v_s = G^p v_0. $ It is however easy to see that  
$ \Q_{+} v_s \neq 0 $ for both cases. Thus there is no singular vector.  
\item 
deg $ u_0 = (0,1) :$  One readily see that $ \X_- u_0 = -m v_0. $ Thus there is no singular vector for $ m \neq 0. $ 
Noting that $ P u_0 = H u_0 = 0, $ one can apply Proposition \ref{prop1} for  $ m = 0 $ and obtain 
$ v_s = G^p u_0 \ (p \in {\mathbb Z}_{\geq 0}) $ as candidates for singular vectors. 
It is easy to verify that $ \Q_{\pm} v_s = 0 $ so that $ v_s $ are indeed singular vectors. 
\item 
deg $ u_0 = (1, -1) : $ One readily see that $ \X_- u_0 = P u_0 = H u_0 = 0. $ 
However, there is no singular vector in this case because of the same reason as the case i). 
\item
deg $ u_0 = (1,1) : $ The condition (\ref{condition1}) yields $ \X_- u_0 = - (\alpha + m \beta) G v_0. $
If $ m \neq 0, $ then $ \alpha = - m \beta $ and $ P u_0 = H u_0 = 0. $ A candidate of the singular vector 
is given by $ v_s = (G^2 - 2mK)^p u_0,$ but one readily see that $ \Q_- v_s \neq 0. $ Thus there is no 
singular vector for $ m \neq 0. $ If $ m = 0, $ then the condition (\ref{condition1}) yields $ \alpha = 0. $ 
Therefore $ u_0 $ is reduced to the one of the case ii). 
\item 
deg $ u_0 = (1,0) : $ The condition (\ref{condition1}) yields $ \X_- u_0 =\alpha m \S_- v_0. $ 
If $ m \neq 0, $ then $ \alpha = 0 $ so that $ u_0 $ is reduced to the one of the case i). 
If $ m = 0 $ then $ \alpha, \beta $ are arbitrary. Furthermore one verifies that $ P u_0 = H u_0 = 0. $ 
Setting $ \beta = 0, $ it is verified that $ v_s = G^p \S_- \X_+ v_0 \ (p \in {\mathbb Z}_{\geq 0}) $ satisfy 
$ \Q_{\pm} v_s = 0 $ if $ r = d - p - 1. $ Therefore they are singular vectors. 
\item 
deg $ u_0 = (1,2) : $ It is immediate to see that $ \X_- u_0 = (m \S_+ - G \X_+) v_0 \neq 0. $ 
Thus no singular vectors. 
\item 
deg $ u_0 = (2,0) : $ The vector $ u_0 $ is given by
\beq
  u_0 = (\alpha \, G \S_- \X_+ + \beta \S_+ \S_- + \gamma \, G^2 + \delta K) v_0.  \label{u0-20}
\eeq
It follows that $ \X_- u_0 = ( \alpha m - \beta) G \S_- v_0. $ 
If $ m \neq 0 $ then $ \beta = \alpha m. $ Furthermore both $P$ and $ H$ does not annihilate $ u_0:$
\beq
  P u_0 = \{ -2m ( \alpha-\gamma) + \delta \} G v_0, \quad
  H u_0 = \{ m (d - r-1)\alpha + m \gamma - d \delta \} v_0. 
  \label{PHaction20}
\eeq
The singular vectors may have the form of
\beq
   v_s = \sum_{\ell} a_{\ell} G^{n-2\ell} K^{\ell} u_0, \quad n = k + 2 \ell 
   \label{vs20}
\eeq
The condition $ \Q_{\pm} v_s = 0 $ yields the following four recurrence relations:
\bea
  & & \{ (d-r-n+2\ell) \alpha + k \gamma \} a_{\ell+1} + (n-2\ell) \delta a_{\ell} = 0,
  \label{rec1} \\
  & & (\ell+1) \gamma a_{\ell+1} + \{ (-d+r + n-2\ell) m \alpha + (\ell+1) \delta \} a_{\ell} = 0,
  \label{rec2} \\
  & & (\ell+1) a_{\ell+1} + (n-2\ell) m a_{\ell} = 0,
  \label{rec3} \\
  & & (\ell+1) \gamma a_{\ell+1} + \{ (d+r-2\ell-1) m \alpha + (\ell+1) \delta \} a_{\ell} = 0.
  \label{rec4} 
\eea
The condition $ P v_s = 0 $ yields the recurrence relation below in addition to (\ref{rec3}):
\beq
  \{ -2m \alpha + (n-2\ell) m \gamma + ( \ell+2) \delta \} a_{\ell+1} + (n-2\ell) m \delta a_{\ell} + (\ell+2) \gamma a_{\ell+2} = 0.
  \label{rec5}
\eeq
In the derivation of (\ref{rec1})-(\ref{rec5}) we used $ \alpha \neq 0, $ since if $ \alpha = 0 $ then $ u_0 $ is 
reduced to the one of the case i). Substitution of (\ref{rec3}) into others and after some algebra we obtain
\beq
  d = \frac{1}{2}(n+1), \qquad \gamma = \frac{d+r+1}{2d+1} \alpha, \qquad \delta = 2m (\alpha-\gamma).
  \label{d-g-d-20}
\eeq
The relation (\ref{rec3}) is identical to the non-super case \cite{DDM}. By solving it we obtain for $ n = 2p $ the expression 
$ f(G,K) = (G^2 - 2mK)^p. $ We  thus obtain a singular vector for $ d = p+1/2 \ (p \in {\mathbb N}) $ and arbitrary value 
of $r.$ We remark that the $ u_0 $ with the condition (\ref{d-g-d-20}) is annihilated by $ P $ and $ H $ but not by $ \Q_{\pm}. $ 

 \quad If $ m = 0 $ we see that the condition (\ref{condition1}) is equivalent to $ \beta = 0. $ The vector (\ref{u0-20}) with $ \beta = 0 $ 
is not annihilated by both $ P $ and $ H. $ We thus use the expression (\ref{vs20}) again. It is then easy to see that 
$ P v_s = 0 $ yields $ \alpha = 0. $ Thus our $ u_0 $ is reduced to the one on the case i). 
\item 
deg $ u_0 = (2,1) : $ The vector $ u_0 $ is given by
\beq
  u_0 = ( \alpha \S_+ \S_- \X_+ + \beta G \S_+ + \gamma\, G^2 \X_+ + \delta K \X_+ ) v_0. \label{u0-21}
\eeq
It is immediate to verify that 
\[
  \X_- u_0 = - \alpha (G \S_- \X_+ + m \S_+ \S_- ) v_0 - (\beta + m \gamma ) G^2 v_0 - m  \delta K v_0.
\]
It follows that if $ m \neq  0 $ then $ \alpha = \delta = 0 $ thus $ u_0 $ is reduced to the case iv). 
If $ m = 0$ then $ \alpha = \beta = 0. $ Thus our $ u_0 $ is reduced to the case ii). 
\end{enumerate}
Now our investigation is summarized in the next proposition:
\begin{prop} \label{prop4}
 The singular vectors of the Verma module $ V^{d,r} $ over \SSch{2} are given as follows: 
 \begin{enumerate}
    \item For $ m \neq 0, $ a singular vector exist for $ d = p+1/2 \ ( p \in {\mathbb N}) $ and arbitrary value of $r:$
    \bea
        & &  v_s^p = (G^2-2mK)^p u_0, \nn \\
        & & u_0 = (G \S_- \X_+ + m \S_+ \S_- + 2mK) v_0 + \frac{d+r+1}{2d+1} (G^2 - 2mK) v_0.
        \label{SVN2}
    \eea
    \item For $ m = 0, $ infinitely many singular vector exist for each value of $d$ and $r:$
    \beq
      v_s^p = G^p \X_+ v_0, \quad p \in {\mathbb Z}_{\geq 0}.  \label{SVN2m0}
    \eeq
    In addition to this, if $ r = d-p-1 \ (p \in {\mathbb Z}_{\geq 0}) $ then infinitely many extra singular vectors 
    exits:
    \beq
      \tilde{v}_s^p = G^p \S_- \X_+ v_0.  \label{SVN2m00}
    \eeq
 \end{enumerate}
\end{prop}
\subsection{Irreducible modules}

  In this subsection we examine the reducibility of the Verma module $ V^{d,r}. $ 
We first consider the case of $ m \neq 0. $ 
If $ d \neq p + 1/2 $ for $ p \in {\mathbb Z}_{\geq 0},$ there is no singular vector in 
$ V^{d,r}. $ Thus the Verma module is irreducible. 
For $ d = p + 1/2 $ there is a singular vector (\ref{SVN2}). 
The submodule $ I^{d,r} = U( {\mathfrak s}(1/2)^+) \otimes v_s^p $ is invariant 
under the action of \SSch{2} and $ I^{d,r} \simeq V^{d',r} $ with $ d' = - p - 3/2. $  
From the same discussion as section \ref{IMN1} it can be seen that $ I^{d,r} $ does not 
have singular vectors. Search for the singular vectors in the factor module $ V^{d,r} / I^{d,r} $ 
can be carried out by the same way as section \ref{IMN1}. 
After some work one may conclude that no singular vectors 
are found in $ V^{d,r} / I^{d,r}. $ Thus the module is irreducible. 

  Next we study the case of $ m = 0. $  Since all $ V^{d,r} $ contain infinitely many singular vectors, 
they are reducible. One see from (\ref{SVN2m0}) and (\ref{SVN2m00}) that $ v_s^p = G^p v_s^0, \ 
\tilde{v}_s^p = G^p \S_- v_s^0. $ 
Hence, in order to find the irreducible modules, we need only consider the factor module
\beq
   V^{d,r} / I^0, \qquad I^0 = U( {\mathfrak s}(1/2)^+) \otimes v_s^0. \label{V/I0}
\eeq
The lowest weight vector $ w_0 $ of the factor module $ V^{d,r} / I^0 $ satisfies the condition
\beq
   \Q_{\pm} w_0 = \X_{\pm} w_0 = P w_0 = 0, \label{LWconN2}
\eeq
and the basis of $ V^{d,r} / I^0 $ is given by
\beq
     G^k K^{\ell} \S_+^a \S_-^b w_0, \quad k, \ell \in {\mathbb Z}_{\geq 0}, \ a, b \in \{ 0, 1 \}
     \label{besisw0}
\eeq
The singular vectors may have the same form as (\ref{SVassumpS2}) provided that $ u_0 $ containing 
only $ \S_{\pm} $ as fermionic generators. All possible  $ u_0 $ are listed in Table~2. 
Each $ u_0 $ is specified by the pair of integers $ (a+b, a-b). $ 
\begin{table}[h]
     \renewcommand{\arraystretch}{1.2}
     \renewcommand{\tabcolsep}{3mm}
 \begin{center}
  \begin{tabular}{ccccc}
    \hline
       & $a$ &  $b$  & $(a+b, a-b)$  &  $u_0$  \\
    \hline\hline
      i) &  0  & 0   & (0,0) &  $w_0$  \\
      ii) & 0 & 1 & (1,-1) & $ \S_- w_0 $ \\
      iii) & 1 & 0 & (1,1) & $ \S_+ w_0 $ \\
      iv) & 1 & 1 & (2,0) & $ (\alpha \S_+ \S_- + \beta G^2 + \gamma K) w_0$ \\
    \hline
  \end{tabular}
 \end{center}
  \caption{List of $u_0$ for $m=0$}
\end{table}
The necessary condition for the singular vectors in this case is 
\beq
  \X_{\pm} u_0 = 0,  \label{condition2}
\eeq
since $ \X_{\pm} $ commute with $ G $ and $ K. $ 
It is immediate to see that the cases ii) iii) do not satisfy the condition. 
For the $ u_0 $ of the case iv) one has $ \X_{\pm} u_0 = \pm \alpha \, G \S_{\pm} w_0 = 0 $ which 
means $ \alpha = 0. $ Thus the case iv) is reduced to the case i). 
Clearly, the vector $ u_0 $ of the case i) is a lowest weight vector of the 
non-super \Sch{1}. Hence, from Proposition \ref{prop1} a candidate of the singular vector is 
given by $ w_s^p = G^p w_0 \ (p \in {\mathbb N}). $ 
It is also easy to see that these $ w_s^p $ are annihilated by $ \Q_{\pm} $  and $ P. $ Thus these $ w_s^p $ are 
singular vectors in $ V^{d,r} / I^0, $ namely, the module $ V^{d,r} / I^0 $ is reducible. 

 Because we have the relation $ w_s^p = G^{p-1} w_s^1,$ the factor module we need to investigate next is 
given by
\beq
  L^{d,r} \equiv ( V^{d,r} / I^0) / {\cal I}^1, \qquad {\cal I}^1 = U( {\mathfrak s}(1/2)^+) \otimes w_s^1. 
  \label{V/I0/I1}
\eeq
The lowest weight vector $ z_0 $ of $ L^{d,r} $ satisfies the condition
\beq
  \Q_{\pm} z_0 = \X_{\pm} z_0 = P z_0 = G z_0 = 0. \label{LWconN2-2}
\eeq
The basis of $ L^{d,r} $ is 
\beq
   K^{\ell} \S_+ ^a \S_-^b z_0, \quad \ell \in {\mathbb Z}_{\geq 0}, \ a, b \in \{ 0, 1 \}
   \label{basisz0}
\eeq
We remark that $ P, G, M $ and $ \X_{\pm} $ are represented trivially on the basis, 
that is, the module $ L^{d,r} $ is also the one of the subalgebra $ osp(2/2). $ 
A singular vector in $ L^{d,r} $ may have the form
\beq
   z_s^{\ell} =  K^{\ell} u_0, \label{SV-L}
\eeq
where $ u_0 $ is a homogeneous element of $ L^{d,r} $ containing each $ \S_{\pm} $ at most once. 
Thus the vector $ u_0 $ is specified by the pair of integer $ (a+b, a-b). $ This means that possible 
$ u_0 $ is identical to the ones given in Table~2 provided that $ w_0 $ is replaced with $ z_0 $ and that 
the terms having $G^2$ disappears from $ u_0 $ for the case iv). 
Because  $ P, G , \X_{\pm} $ have a trivial representation, $ z_s^{\ell} $ is a singular vector 
if $ \Q_{\pm} z_s^{\ell} = 0. $ We examine this condition for possible $ u_0 $ according to Table~2.  

\medskip\noindent
i) $ u_0 = z_0 : $ For this case we have $ \Q_{\pm} z_s^{\ell} = \ell K^{\ell-1} \S_{\pm} z_0. $ 
Thus the condition requires $ \ell = 0 $ which is trivial. 
Thus there is no singular vector for this case.

\medskip\noindent
ii) $ u_0 = \S_- z_0 : $ One verifies easily that $ \Q_- z_s^{\ell} = 0 $ and
\[
  \Q_+ z_s^{\ell} = \{ (d-r) k^{\ell} + \ell K^{\ell-1} \S_+ \S_- \} z_0.
\] 
It follows that $ z_s^0 = \S_- z_0 $ is a singular vector if $ d = r. $

\medskip\noindent
iii) $ u = \S_+ z_0 : $ One readily see that $ \Q_+ z_s^{\ell} = 0 $ and
\[
   \Q_- z_s^{\ell} = \{ (d+r) K^{\ell} + \ell K^{\ell-1} \S_- \S_+ \} z_0.
\]
It follows that $ z_s^0 = \S_+ z_0 $ is a singular vector if $ d = -r. $ 
 
\medskip\noindent
iv) $ u = ( \alpha \S_+ \S_- + \beta K) z_0 : $ The condition yields the following relations:
\bea
  & & -(d-r) \alpha + (\ell+1) \beta = 0, \nn \\
  & & (d+r-2\ell-2) \alpha + (\ell+1) \beta = 0. \nn
\eea
It turns out by solving these relations ($ \alpha \neq 0 $)  that if $ d = \ell+1 $ we have a singular vector:
\beq
  z_s^{\ell} = K^{\ell} \S_+ \S_- z_0 + \frac{d-r}{d} K^{\ell+1} z_0.
  \label{SV-iv}
\eeq 
Therefore $ L^{d,r} $ is irreducible if $ r \neq \pm d $ or $ d $ is not a positive integer. 
On the other hand, if $ r = \pm d $ or $ d = p \in {\mathbb N}, $ then $ L^{d,r} $ is reducible. 
We thus need to investigate three factor modules:
\bea
  & & {\cal L}_-^d = L^{d,d} / U({\mathfrak s}(1/2)^+) \otimes \S_- z_0,
  \nn \\[3pt]
  & & {\cal L}_+^d = L^{d,-d} / U({\mathfrak s}(1/2)^+) \otimes \S_+ z_0,
  \nn \\[3pt]
  & & {\cal L}^{p,r} = L^{p,r} / U({\mathfrak s}(1/2)^+) \otimes z_s^{p-1}, \quad
      \mbox{with} \ z_s^{p-1} \ \mbox{in (\ref{SV-iv})}
  \nn
\eea
We denote the lowest weight vector of both $ {\cal L}_{\pm}^d $ by $ \ket{0} $ (this does not cause 
any confusion). Then it is annihilated by $ \Q_{\pm}, \X_{\pm}, P, G $ and $ \S_+ $ for $  {\cal L}_+^d $ 
and by $ \Q_{\pm}, \X_{\pm}, P, G $ and $ \S_- $ for $  {\cal L}_-^d. $
The bases of $  {\cal L}_+^d $ and $ {\cal L}_-^d $ are given by
\[
   K^{\ell} \S_-^a \ket{0}, \qquad K^{\ell} \S_+^a \ket{0}, \quad \ell \in {\mathbb Z}_{\geq 0}, \ a \in \{ 0, 1 \}
\]
respectively. It follows that $  {\cal L}_+^d $ and $ {\cal L}_-^d $ are isomorphic. The isomorphism is 
given by $ \omega_2 $ defined in (\ref{omega2-even}) (\ref{omega2-odd2}) or  $ \sigma_2 $ defined in (\ref{GAO2-S1}). 
We thus examine $  {\cal L}_+^d $ and $ {\cal L}^{p,r}. $ 

  Search for the singular vector in $  {\cal L}_+^d $ is equivalent to find the basis vectors annihilated by $ \Q_{\pm}. $ 
The basis vectors with $ a = 0 $ are annihilated by $ \Q_+ $ and those with $ a = 1 $ are  by $ \Q_-.$ 
While we have the followings:
\beq
  \Q_- K^{\ell} \ket{0} = \ell K^{\ell-1} \S_- \ket{0}, \qquad
  \Q_+ K^{\ell} \S_- \ket{0} = 2 (d - \ell) K^{\ell} \ket{0}.
  \label{QonBasis}
\eeq
The first relation of (\ref{QonBasis}) requires $ \ell = 0 $ for $ a = 0. $ This shows that there is no singular 
vector for $ a = 0. $ It follows from the second relation of  (\ref{QonBasis}) that if $ d = \ell $ then 
$ K^{\ell} \S_- \ket{0} $ is a singular vector. Hence, $  {\cal L}_+^d $ is irreducible if $ d $ is not a non-negative 
integer. For $ d = \ell \in {\mathbb Z}_{\geq 0} $ we consider the factor module $ {\cal L}_+^{\ell} / {\mathscr I}^{\ell} $ 
where $ {\mathscr I}^{\ell} = U({\mathfrak s}(1/2)^+) \otimes K^{\ell} \S_- \ket{0}. $ 
The basis of $ {\cal L}_+^{\ell} / {\mathscr I}^{\ell} $ is given by
\[
  \S_-^a \ket{\tilde{0}}, \ K \S_-^a \ket{\tilde{0}}, \  K^2 \S_-^a \ket{\tilde{0}}, \ \cdots \ 
  K^{\ell-1} \S_-^a \ket{\tilde{0}}, \ K^{\ell} \ket{\tilde{0}}, \quad
  a \in \{ 0, 1 \}
\]
where $ \ket{\tilde{0}} $ is the lowest weight vector of $ {\cal L}_+^{\ell} / {\mathscr I}^{\ell}. $ 
It is easy to see that there is no singular vector in $ {\cal L}_+^{\ell} / {\mathscr I}^{\ell}. $ 
Therefore, $ {\cal L}_+^{\ell} / {\mathscr I}^{\ell} $ is irreducible. 

  Next we study $ {\cal L}^{p,r}. $ 
We denote the lowest weight vector of $ {\cal L}^{p,r} $ by $ \ket{0}. $ Then
\bea
  & & \Q_{\pm} \ket{0} = \X_{\pm} \ket{0} = P \ket{0} = G \ket{0} = 0,
  \nn \\[3pt]
  & & K^p \ket{0} = \frac{p}{p-r} K^{p-1} \S_+ \S_- \ket{0}. \nn
\eea 
It follows that $ K^p \S_{\pm} \ket{0} = K^p \S_+ \S_- \ket{0} = 0. $ Thus the basis of $ {\cal L}^{p,r} $ 
is given by
\beq
  K^{\ell} \S_+^a \S_-^b \ket{0}, \quad \ell < p, \ a, b \in \{ 0, 1 \}
  \label{basisLpr}
\eeq
Singular vectors in $ {\cal L}^{p,r} $ will be obtained by imposing the condition that the vectors are annihilated by $ \Q_{\pm}. $ 
If we carry this out according to the classification same as Table~2 then it turns out that  
the process is almost same as the case of $ L^{d,r}. $ We therefore mentions the results and omit the detail. 
The case i) has no singular vectors. 
The case ii) has one singular vector $ \S_- \ket{0} $ if $ r = d. $ Thus the analysis of this case gives the equivalent 
result to $ {\cal L}_-^d. $ The case iii) also has one singular vector $ \S_+ \ket{0} $ if $ r = -d. $ 
This case is reduced to $ {\cal L}_+^d. $ 
The case i) does not have any singular vectors. 

  Summarizing the results obtained so far, we have proved the following proposition.
\begin{prop}
  The irreducible lowest weight modules over the Lie superalgebra \SSch{2} are classified as follows:
 \begin{enumerate}
   \item $ m \neq 0$
    \begin{itemize}
      \item $ V^{d,r} $ when $ d \neq p + 1/2, \ p \in {\mathbb Z}_{\geq 0}. $ $ dim V^{d,r} = \infty. $
      \item $ V^{d,r} / I^{d,r} $ when $ d = p + 1/2, \ p \in {\mathbb Z}_{\geq 0}. $ $ dim V^{d,r} / I^{d,r} = \infty. $
    \end{itemize}
    For both cases, $r$ takes an arbitrary value. 
   \item $ m = 0 $
    \begin{itemize}
       \item $ L^{d,r} $ when $ r \neq \pm d $ or $ p$ is not a positive integer. dim $ L^{d,r} = \infty. $
       \item $ {\cal L}_+^d \simeq {\cal L}_-^d $ when $ d $ is not a non-negative integer. dim $ {\cal L}_+^d = \infty. $
       \item $ {\cal L}_+^{\ell} / {\mathscr I}^{\ell} $ when $ \ell \in {\mathbb Z}_{\geq 0}. $ 
             dim $ {\cal L}_+^{\ell} / {\mathscr I}^{\ell} = 2\ell +1. $
    \end{itemize}
    The representations for $ m = 0 $ are also irreps of the subalgebra $ osp(2/2). $ 
 \end{enumerate}
\end{prop}

%
%
%
\setcounter{equation}{0}
\section{Vector field realization}
\label{VFR}

In this section we present a vector field realization of \SSch{1} and \SSch{2} which is an 
extension of the standard vector field realization of the Schr\"odinger algebra \Sch{1}. 
The standard vector field realization of \Sch{1} has one time and one space coordinates $ (t,x). $ 
In order to realize \SSch{1} we further introduce two variables $ (\theta, \eta) $ of parity odd. 
One of them is Grassmannian and the another is related to mass eigenvalue:
\beq
 \{ \theta, \theta \} = \{ \theta, \eta \} = 0, \qquad \{ \eta, \eta \} = -m.
 \label{N1-space}
\eeq
With the four variables $ (t,x,\theta,\eta) $ the realization of \SSch{1} is given by
     \bea
       & & K = t ( t \del_t + x \del_x + \theta \del_{\theta}) + \frac{m}{2}x^2 + x \theta \eta - td,
       \nn \\[3pt]
       & & G = t \del_x + mx + \theta \eta,
          \qquad D = 2t \del_{t} + x \del_{x} + \theta \del_{\theta} - d,
       \nn \\[3pt]      
       & & H = \del_{t}, \qquad P = \del_x,\qquad M = m, \qquad \Q = -\theta \del_t + \del_{\theta}, 
       \nn \\[3pt]
       & & \S = -\theta ( t \del_t + x \del_x) + t \del_{\theta} + x \eta + \theta d,
       \qquad
       \X = - \theta \del_{x} + \eta,
       \label{N1Vec}
     \eea
where $d$ is the conformal weight. 
To realize \SSch{2} we introduce three Grassmann variables $(\theta,\phi,\rho). $ 
The realization is given as follows:
        \bea
          & & K = t( t \del_t + x \del_x + \theta \del_{\theta} + \phi \del_{\phi})
              + \theta \phi \rho \del_{\rho} - mx \theta \rho 
              + \frac{m}{2}x^2 + x \phi \del_{\rho} -t d,
          \nn \\
          & & 
            G = t \del_x + m(x- \theta \rho) + \phi \del_{\rho}, \qquad P = \del_{x},
          \nn \\[3pt]
          & & 
             D = 2 t\del_t + x \del_x + \theta \del_{\theta} + \phi \del_{\phi} - d,
          \nn \\[3pt]
          & &
             H = \del_{t}, \qquad R = - \theta \del_{\theta} + \phi \del_{\phi} + \rho \del_{\rho}, \qquad M = m,
          \nn \\[3pt]
          & & 
            \Q_+ = - \phi \del_{t} + \del_{\theta}, \qquad\ \;
            \Q_- = - \theta \del_{t} + \del_{\phi}
          \nn \\
          & & 
            \S_+ = \phi(-t \del_t - x \del_x - \theta \del_{\theta} + \rho \del_{\rho} ) + t \del_{\theta}
                 - m x \rho + \phi d,
          \nn \\[3pt]
          & & 
            \S_- = \theta (-t \del_t - x \del_x - \phi \del_{\phi} - \rho \del_{\rho} ) + t \del_{\phi}
                 + x \del_{\rho} + \theta d,
          \nn \\[3pt]
          & & 
            \X_+ = - \phi \del_{x} - m \rho, \qquad \X_- = -\theta \del_{x} + \del_{\rho}. 
          \label{N2Vec}
        \eea

  We have introduced two Clifford-like elements $\chi $ and $ \eta $ for \SSch{1}. 
We remark that they are realized by a single Grassmann number $\phi $ as follows:
\beq
  \chi = \sqrt{ \frac{m}{2} } ( \phi + \del_{\phi} ), \qquad \eta  = \sqrt{ \frac{m}{2} } ( \phi - \del_{\phi} ).
  \label{chi-eta}
\eeq 

  The vector field realizations given above together with singular vectors may give 
invariant partial differential equations of super Schr\"odinger algebras. 
This is a supersymmetric extension of the result for semisimple Lie algebras developed in \cite{Dob88,Kos}. 
The invariant equations obtained by this procedure for \Sch{n} have been obtained in \cite{DDM,Mur,ADD,ADDS}.

%
%
%
\setcounter{equation}{0}
\section{Concluding remarks}
\label{ConRem}

  We investigated the lowest weight modules (Verma modules) over \SSch{1} and \SSch{2}. 
Explicit expression of the singular vectors were derived and reducibility of the Verma modules 
has been studied. This led us to classify irreducible modules over \SSch{1} and \SSch{2}. 

  We comment on a bilinear form analogous to  the Shapovalov form \cite{Sha} of the semisimple Lie algebra. 
Let $ {\mathfrak g} = $ \SSch{1} or \SSch{2} and $ V $ be a Verma module over $ {\mathfrak g}. $ 
The lowest weight vector of $ V $ is denoted by $ v_0 $ as usual. 
We define the bilinear form $ (\ ,\ ) : V \otimes V \rightarrow {\mathbb C} $  
by the relations:
\beq
  (X v_0, Y v_0) = (v_0, \omega_1(X) Y v_0), \qquad (v_0, v_0) = 1, \quad X, Y \in U({\mathfrak g})
  \label{ShaForm}
\eeq
If $ v_m, v_n \in V $ have different weight, then they are orthogonal with respect to this form:
\beq
  (v_m, v_n) = 0.  \label{Ortho}
\eeq 
To see this, suppose that the weights of $ v_m, v_n $ are $m$ and $ n, $ respectively. 
Then
\[
  (D v_m, v_n) = m (v_m, v_n).
\]
The left hand side has alternate way of computation:
\[
  (D v_m, v_n) = (v_m, \omega_1(D) v_n) = n (v_m, v_n).
\]
Since $ m \neq n $ we obtain (\ref{Ortho}). 
It follows that a singular vector of ${\mathfrak g}$ is orthogonal to any other vectors in $V.$ 
This is the same property as semisimple case. 
Thus one may analyze the reducibility of the Verma modules also via the bilinear form. 

  We provided a vector field realization of \SSch{1} and \SSch{2} in section \ref{VFR}. 
It will open a way to physical applications of the super Schr\"odinger algebras. 
One example we will do as a future work is a supersymmetric extension of the group theoretical 
approach to nonrelativistic holography discussed in \cite{AV}. 
Another important future work is the classification of irreducible modules for the super 
Schr\"odinger algebras of higher dimensional 
spacetime. Especially, the most physical $(1+3)$ dimensional spacetime is of importance. 
We have investigated $ {\cal N} = 1, 2 $ in the present paper. This is because physical 
applications in nonrelativistic setting are known for small values of $ {\cal N}. $ 
Of course, this does not mean the super Schr\"odinger algebras for large values of ${\cal N} $ 
are useless. Analysis of irreducible representations for large $ \cal N $ is also an 
interesting problem.

%
%
%

\end{document}